\documentclass[twocolumn,
aps,
prb,
showpacs,
amsmath,
amssymb,
floatfix,
reprint,nofootinbib, superscriptaddress]{revtex4-1}
\pdfoutput=1

\usepackage{comment}  
\usepackage[T1,T2A]{fontenc}
\usepackage[utf8]{inputenc}
\usepackage{graphicx}        
\usepackage{bm}         
\usepackage{siunitx}
\usepackage{color}
\definecolor{darkblue}{rgb}{0.0,0.0,0.3}
\definecolor{goodblue}{rgb}{0.0,0.0,0.6}
\usepackage{float}
\usepackage{hyperref}
\usepackage{xfrac}
\usepackage{placeins}
\hypersetup{
	colorlinks   = true, 
	urlcolor     = darkblue, 
	linkcolor    = darkblue, 
	citecolor   = darkblue 
}

\bibpunct{[}{]}{,}{n}{}{}
\makeatletter
\def\NAT@def@citea{\def\@citea{\NAT@separator}}
\makeatother

\begin{document}
\title{Optical Enhancement of Superconductivity via Targeted Destruction of Charge Density Waves} 

\author{Hossein Dehghani}
\altaffiliation{Correspondence to \href{mailto:hdehghan@umd.edu}{hdehghan@umd.edu}}
\affiliation{Joint Quantum Institute, College Park, 20742 MD, USA}
\affiliation{The Institute for Research in Electronics and Applied Physics, University of Maryland, College Park, 20742 MD, USA}

\author{Zachary M. Raines\footnotemark[1]}
\affiliation{Joint Quantum Institute, College Park, 20742 MD, USA}
\affiliation{Condensed Matter Theory Center, University of Maryland, College Park, 20742 MD, USA}

\author{Victor M. Galitski}
\affiliation{Joint Quantum Institute, College Park, 20742 MD, USA}
\affiliation{Condensed Matter Theory Center, University of Maryland, College Park, 20742 MD, USA}

\author{Mohammad Hafezi}
\affiliation{Joint Quantum Institute, College Park, 20742 MD, USA}
\affiliation{The Institute for Research in Electronics and Applied Physics, University of Maryland, College Park, 20742 MD, USA}

\date{\today}
\newcommand{\hsigma}{\ensuremath{\hat{\sigma}}} 
\newcommand{\tr}{\ensuremath{\mathrm{tr}}}
\newcommand{\eff}{\ensuremath{\mathrm{eff}}}

\newcommand{\omegak}{\ensuremath{\omega_k}}
\newcommand{\halvedBZ}{\ensuremath{\widetilde{\mathrm{BZ}}}}
\newcommand{\bpartial}{\boldsymbol{\partial}}

\newcommand{\bD}{\mathbf{D}}
\newcommand{\bQ}{\mathbf{Q}}
\newcommand{\br}{\mathbf{r}}

\newcommand{\bk}{\mathbf{k}}
\newcommand{\bx}{\mathbf{x}}
\newcommand{\by}{\mathbf{y}}
\newcommand{\thetadelta}{\theta_{\delta}}

\newcommand{\bK}{\mathbf{K}}
\newcommand{\bS}{\mathbf{S}}
\newcommand{\bq}{\mathbf{q}}
\newcommand{\bLambda}{\mathbf{\Lambda}}
\newcommand{\plambda}{k_{\Lambda}}
\newcommand{\dlambda}{d_{\Lambda}}
\newcommand{\bp}{\mathbf{p}}
\newcommand{\bA}{\mathbf{A}}
\newcommand{\bv}{\mathbf{v}}
\newcommand{\mf}{\ensuremath{\mathrm{MF}}}
\newcommand{\hh}{\ensuremath{\hat{h}}}

\begin{abstract}
It has been experimentally established that the occurrence of charge density waves is a common feature of various under-doped cuprate superconducting compounds. The observed states, which are often found in the form of bond density waves (BDW), often occur in a temperature regime immediately above the superconducting transition temperature. Motivated by recent optical experiments on superconducting materials, where it has been shown that optical irradiation can transiently improve the superconducting features, here, we propose a new approach for the enhancement of superconductivity by the targeted destruction of the BDW order. Since BDW states are usually found in competition with superconductivity, suppression of the BDW order enhances the tendency of electrons to form Cooper pairs after reaching a steady-state. By investigating the optical coupling of gapless, collective fluctuations of the BDW modes, we argue that the resonant excitation of these modes can melt the underlying BDW order parameter. We propose an experimental setup to implement such an optical coupling using 2D plasmon-polariton hybrid systems.       

\end{abstract}


\maketitle 
\section{Overview}
Cuprate high temperature superconductors, which are made up of quasi two-dimensional layers of copper-oxide materials, have been a subject of intense interest in quantum condensed matter physics since their experimental discovery in the 1980s~\cite{bednorz1986possible} (see \cite{keimer2015quantum, lee2006doping} for a review of the properties of these materials).
A major feature of these systems is the onset of variety of orders at low temperatures ~\cite{kaminski2002spontaneous, kohsaka2007intrinsic, lawler2010intra, parker2010fluctuating, fujita2014direct}. 
More specifically, there is now mounting experimental evidence that at low temperatures in the underdoped region spontaneous breaking of crystalline symmetries and translational symmetries result in charge density wave states, characterized by the nontrivial modulation of charge density. Since this modulation is centered on the bonds of the square lattice rather than on the sites, these modes are often referred to as ``bond density waves.'' These states have been detected universally in all cuprate families, via various experimental techniques such as scanning tunneling microscopy (STM) and X-ray scattering \cite{wise2008charge, achkar2012distinct,torchinsky2013fluctuating,comin2014charge, comin2015symmetry,comin2015broken,hamidian2016atomic,mesaros2016commensurate}. While the debate is still ongoing regarding the microscopic origin of the BDW state, temperature-dependent measurements in the presence of magnetic fields and external pressure suggest that the BDW order and superconductivity (SC) compete over a wide doping region \cite{gabovich2010competition, chang2012direct,ghiringhelli2012long, achkar2012distinct, wu2013emergence, blackburn2013x, huecker2014competing, blanco2014resonant, wu2015incipient, wang2018fragility, loret2019intimate}.

This competition between the BDW and SC orders can be explained phenomenologically by considering the Ginzburg-Landau theory of the SC order parameter $\Delta$, and BDW order parameter $\Phi$.
From symmetry considerations, one can write a Ginzburg-Landau theory for the total free energy as a function of the temperature $T$
\cite{landau1987statistical, kivelson2002competing}, 
\begin{align}
    \mathcal{F}[\Delta, \Phi] &= a_\Delta(T)|\Delta|^2 + b_\Delta(T) |\Delta|^4 + a_\Phi (T) |\Phi|^2 \notag \\
    & + b_\Phi (T)|\Phi|^4 + C(T) |\Delta|^2|\Phi|^2, \label{freeEnergy}
\end{align}
where the exact values of the coefficients and their temperature-dependence are determined from the electronic theory in terms of the polarization functions \cite{raines2015enhancement}. Also for simplicity, we have suppressed gradient terms.

We note that the coefficient of the last term in the free energy, which couples the SC and BDW orders, is always positive $C(T)> 0$.
Therefore, any decrease in the BDW order $\Phi$ energetically favors the formation of the superconducting order $\Delta$.
Consequently, we can imagine a protocol where we start above the superconducting critical temperature $T_c$, where only the BDW order is present.
Then, by driving the system in such a way as to suppress the BDW order we can enhance the tendency toward superconductivity.
Upon establishing a steady-state, we therefore expect that the superconducting order will be greater than its equilibrium value, which in the case of $T > T_c$ is 0.

Experimentally, this competition between charge density waves and superconductivity has been used to achieve higher values of $T_c$ in out-of-equilibrium settings.
In particular, it has been shown, in a number of experiments, that irradiation with THz laser fields for a duration of hundreds of femtoseconds transiently enhances signatures of electron-electron pairing in cuprates \cite{fausti2011light, kaiser2014optically,  hu2014optically, mankowsky2014nonlinear, casandruc2015wavelength, nicoletti2016nonlinear}.
Theoretically, this effect has often been attributed to the resulting lattice deformations of the Cu-O bonds, which melt the existing charge density wave orders \cite{denny2015proposed, raines2015enhancement,knap2016dynamical, patel2016light, sentef2016theory, coulthard2017enhancement, wang2018light, sun2019transient, niwa2019light}. Alternatively, in other models, it has been shown that a combination of the oscillatory behaviour between superconductivity and charge density wave \cite{bittner2019possible} can result in the dynamical enhancement of superconductivity, when other initially vanishing superconducting pairings are incorporated \cite{sentef2017theory} (see \cite{cavalleri2018photo} for a recent review of the field). 

\begin{figure}
\includegraphics[height=4.2cm,keepaspectratio]{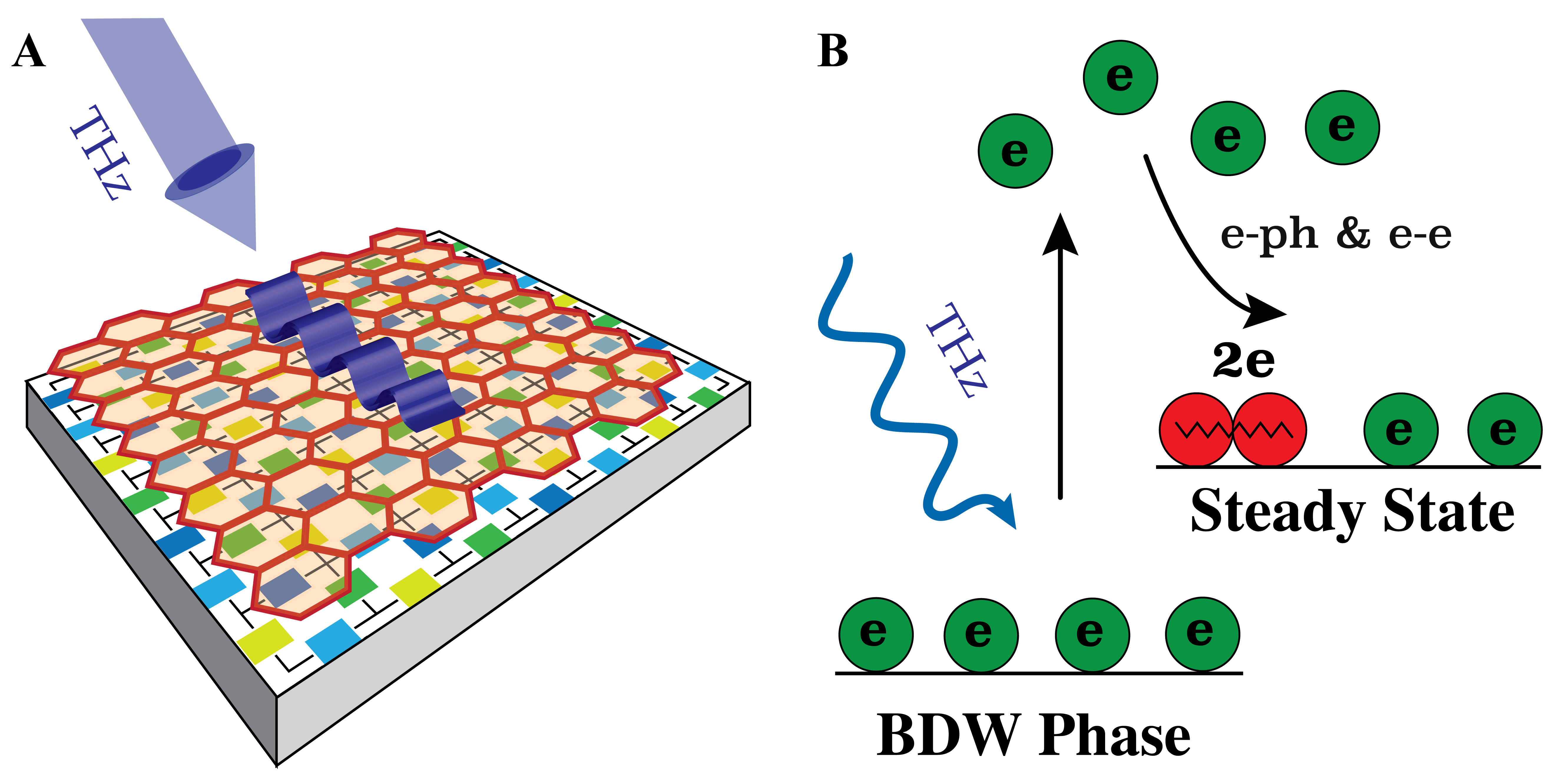}
\caption{
(A) Experimental scheme for generating surface plasmon-polaritons, using a 2D high carrier conductor such as doped graphene, on top of a cuprate superconductor. Irradiation with THz laser fields creates surface plasmon-polariton waves propagating along the conductor and penetrating inside adjacent $\mathrm{CuO_2}$ planes with a bond density wave condensate. 
The color pattern on the $\mathrm{CuO_2}$ plane displays the real part of the charge modulation $P_{ij}$ on the center of the bonds in a bond density phase with (for purposes of illustration) a diagonal wavevector $\bQ = (\pi/2, \pi/2)$.
(B) Schematic depiction of processes after irradiation with light. Absorption of photons by the electrons in the bond density ground state excites these electrons to higher energy states. After the inclusion of electron-electron and electron-phonon scattering processes, the effective exhaustion of the bond density wave channel enhances the formation of Cooper pairs in the non-equilibrium steady-state of the system. 
}\label{SurfacePlasmon}
\end{figure}

Here, inspired by these experiments, we propose a new optical approach to \emph{directly} melt competing electronic modes in a targeted way (Fig.~\ref{SurfacePlasmon}(a)). In contrast to the current experimental methods, which indirectly weaken the BDW state by stimulating phononic vibrations \cite{fausti2011light, kaiser2014optically,  hu2014optically, mankowsky2014nonlinear, nicoletti2016nonlinear}, our proposal is focused on the efficient  optical destruction of the BDW order via optically exciting its collective phase modes. 
The associated excitations are the phase (phasons) and amplitude collective modes of the BDW order parameter. 
In this work, we focus on the phase mode of the BDW order, since it has a gapless spectrum (being the Goldstone mode of the BDW phase) and can also be optically driven at low energies.
Specifically, since interactions within the copper-oxygen planes are believed to play  the dominant role in superconductivity in cuprates, we study the electromagnetic response of the phason mode within a single quasi-2D plane.  
We argue that by exciting phase modes inside these planes, their fluctuations will deplete the BDW order, and therefore, provide an environment more conducive to the formation of non-equilibrium superconductivity.
The interplay between the drive and inherent relaxation mechanisms, as depicted in Fig.~\ref{SurfacePlasmon}(b), then allows for the existence of out-of-equilibrium steady-state superconductor.

To this end, we derive a low energy effective theory for the phason mode of the BDW order from the ``hot-spot model'' of cuprate superconductors~\cite{abanov2000spin, metlitski2010quantum}. This is a minimal theoretical model, built upon the experimental facts regarding the competing orders in cuprates, which does not rely on  a particular microscopic mechanism underlying the orders. The model allows us to obtain the optical response of the BDW collective modes, which determines the properties of their optical coupling. From this result, the optical power absorbed by these modes is calculated and used to obtain a phenomenological estimate for the melting rate of the BDW order parameter. 

More importantly, we derive the momentum-energy matching condition between the matter and optical modes. Experimentally, this condition cannot be realized via conventional optical methods since the speed of light is nearly two orders of magnitude higher than the Fermi velocity of electrons in solids, which determines the propagation speed of BDW collective modes. 
Nevertheless, by hybridizing photons with surface plasmons and creating hybrid modes known as ``surface plasmon-polaritons'' (SPP) the propagation of light can be confined along a metal-dielectric interface \cite{basov2016polaritons}, which shrinks the effective wavelength of light. 
Concretely, placing a 2D (semi-)metal sheet, e.g. graphene, silver, beryllium, in contact with a dielectric slab, the optical properties of the metallic system are modified so that the effective wavelength of photons traveling along the metal-insulator interface $\lambda_{\mathrm{sp}}$, is shrunk by the surface plasmon-polariton confinement ratio $\lambda_{\mathrm{sp}}/\lambda_o$ where $\lambda_o$ labels the wavelength of photons in vacuum~\footnote{In general, to create surface plasmon-polaritons, one sandwiches a metal between two dielectric slabs, one of which can be air, as depicted in Fig.~\ref{SurfacePlasmon}.}.
Thus, by creating a heterostructure of a cuprate superconductor and an appropriate metal-dielectric SPP, as is schematically illustrated in Fig.~\ref{SurfacePlasmon}(a), the energy and momentum mismatch can be remedied in a relatively wide range of frequencies.

The pumping of the system out of the BDW manifold is counter balanced by transition elements from the excited state to the SC state and back to the BDW state as schematically depicted in Fig.~\ref{SurfacePlasmon}(b).
These transition rates together form a non-linear rate equation for the density matrix, whose solutions determine the steady-state behavior of the system.
Even after placing a surface plasmon-polariton material on top of our superconductor, there are no relevant transitions out of the SC state with which our pumping scheme is resonant.
As superconductivity is the closest subleading instability, we expect an enhancement of superconductivity to naturally follow.

In the rest of this work, we focus on the optical melting of the BDW order and the resultant suppression.

\section{Model and Results}

\subsection{Model} 
The BDW state is described by an incommensurate bond density wave order where the modulations of the charge are, for a single band model, located on the Cu-Cu bonds, or  in models with more bands on the oxygen sites, rather than the Cu atoms \cite{sau2014mean}. Denoting the BDW ordering wavevector by $\bQ$, the BDW order is characterized by a charge expectation value $\Phi P_{ij}$ on the links of the lattice, with the form-factor
\begin{align}
    P_{ij} = \frac{1}{V} e^{i\bQ\cdot(\br_i + \br_j)/2} \sum_{\bk}e^{i\bk\cdot(\br_i -\br_j)}P_{\bQ}(\bk),
\end{align}
where $V$ denotes the volume of our system. 
This instability can be obtained from consideration of the phenomenological t-J-V model ~\cite{sachdev2013bond,sau2014mean,raines2015enhancement} whose Hamiltonian is
\begin{equation}
    H_{\mathrm{t-J-V}} = \sum_{i,j;\alpha}t_{ij} c^{\dagger}_{i, \sigma}c_{j,\alpha} + \frac{1}{2}\sum_{\langle i,j \rangle}J\bS_{i}\cdot \bS_{j} + \frac{1}{2}\sum_{\langle i,j \rangle }Vn_i n_j,
\end{equation}
where the spin and density operators at site $i$ are given by $\bS_i = \sum_{\alpha \beta} c^{\dagger}_{i, \alpha} \hat{\tau}_{\sigma\sigma'}c_{i, \sigma}$, and $n_{i} = \sum_{\sigma} c^{\dagger}_{i,\sigma} c_{j, \sigma}$, respectively. Here, $\hat{\tau}$ labels the spin operator, and Greek indices in the summation run over $\sigma, \sigma' = \{\uparrow, \downarrow \}$. The hopping amplitude $t_{ij}$, usually includes up to the third nearest neighbor hoppings; and $J$, and $V$ denote the nearest-neighbor spin-spin and density-density interaction strengths, respectively. 
In terms of the the electronic operators, the BDW state is characterized by the nonzero expectation value $\Phi P_{ij} = \sum_\alpha \langle c^\dagger_{i\sigma}c_{j\sigma}\rangle.$

It should be noted that while the BDW wavevector $\bQ$ is usually observed to be axial \cite{ghiringhelli2012long, chang2012direct, achkar2012distinct}, i.e. $\bQ = (Q, 0),$ or $(0, Q)$, in the t-J-V model introduced above, the optimal instability is found to be diagonal $\bQ = (\pm Q, \pm Q)$ \cite{sau2014mean}. Nonetheless, we expect the nature of our results, after slight modifications, can be applied to other forms of bond density waves with different orientations of the ordering.

In this study, we suppose that there is a spontaneous symmetry breaking down to orders with one of these BDW vectors $\bQ = (-Q, Q)$.
The periodicity of modulation is between $3$ to $5$ lattice constants and the angular symmetry of these orders has been found to be predominantly of ``$d$-wave'' ($B_{1g}$) symmetry $P_{\bQ}(\bk) = \cos(k_x)-\cos(k_y)$\cite{sachdev2013bond}.
The spatial profile of this order parameter for a diagonal wavevector is shown in Fig.~\ref{SurfacePlasmon}(a), where a commensurate wave with periodicity of 4 is displayed for purposes of illustration.     
The dynamics of the BDW collective modes emerge from electronic degrees of freedom.

\begin{figure}
\includegraphics[height=6cm,keepaspectratio]{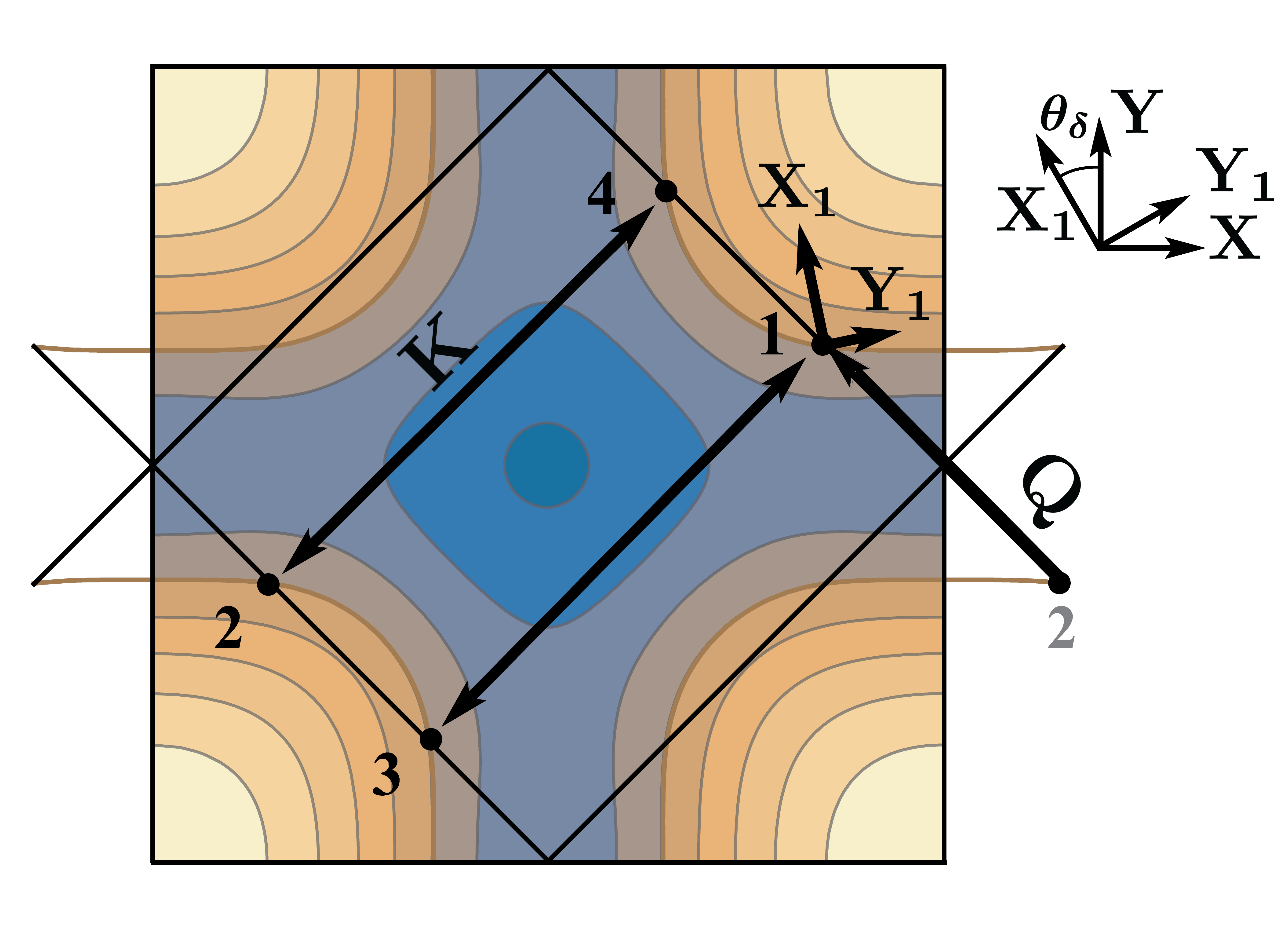}
\caption{
Fermi surface of a square lattice model with up to $3$ nearest neighbor hopping amplitudes. Different colors specify different doping levels. ``Hot-spot'' regions $1,2$, represented by black dots on the Fermi surface, are connected to regions $3,4$, respectively, by the anti-ferromagnetic vector $\bK = (\pi, \pi)$. In the bond density wave phase, electron-hole pairs separated by the momentum $\bQ=(-Q, Q)$, corresponding to the separation of hotspots $1$ and $2$, are condensed. $X_1$ and $Y_1$ are the perpendicular and tangent axes to the Fermi surface at hot-spot $1$, respectively. On the upper right corner the tilt angle $\theta_{\delta}$, defines the angle between the hot-spot axes and the Brillouin zone axes. For the purpose of visual convenience, this angle is exaggerated. 
}\label{FS}
\end{figure}

In this work, we employ the ``hot-spot'' approximation, which allows for considerable improvement in the analytically tractability of the problem.
By restricting to momentum neighborhoods of the so-called ``hot-spots'' on the Fermi surface, the points formed by the intersection of the Fermi surface and the magnetic Brillouin zone, one obtains the low-energy, ``hot-spot'' model \cite{metlitski2010quantum, sachdev2013bond, wang2014charge}.  
These are strongly coupled to each other via anti-ferromagnetic spin fluctations with momentum $\bK =(\pi, \pi)$~\cite{metlitski2010quantum, efetov2013pseudogap}, which are commonly believed to be important for the formation of superconductivity in cuprates~\cite{chubukov2008spin}.
There are $8$ such electron regions, but for solutions which preserve the time reversal symmetry, the hot-spot regions are block-diagonalized into two coupled sets. 
One such set is depicted in Fig.~\ref{FS} labeled by $\psi_{\sigma a}$ where $a = 1,\ldots,4$ and $\sigma = {\uparrow, \downarrow}$.  Furthermore, by imposing $d-$wave symmetry in the Brillouin zone, the hot-spot regions $1,2$ becomes redundant with $3,4$.
Hence, in the following, we only keep one pair of hot-spots in our calculations, which will be combined into the  spinor $\Psi_{k}^{\dagger} = (\psi_{1,k}^{\dagger}, \psi_{2,k}^{\dagger})$.
Nonetheless, in the final results, we also collect the contributions from the condensation of e-h pairs in regions $3,4$, which are simply related to those of regions $1,2$ by a $C_4$ rotation.

Note that as in Fig.~\ref{FS}, the momentum axes in the vicinity of hot-spot $1$ are not exactly aligned with the original $X,Y$ axes in the Brillouin zone. Therefore, we define the $X_1$ and $Y_1$ axes, which are perpendicular and tangent to the Fermi surface, respectively, and are rotated by a tilt angle $\delta$ with respect to the original $X$ and $Y$ axes of the Brillouin zone. More explicitly, for an arbitrary wavevector $\bk = (k_x, k_y)$ its projection along the rotated axes can be expressed in terms of $\theta_{\delta}$ according to
\begin{align}
\begin{pmatrix} k_{x_1} \\ k_{y_1}
\end{pmatrix} = 
\begin{pmatrix}
-\sin(\theta_{\delta}) & \cos(\theta_{\delta})\\ -\cos(\theta_{\delta} )& \sin(\theta_{\delta})
\end{pmatrix} \begin{pmatrix} k_{x} \\ k_{y}
\end{pmatrix}\label{theta}
\end{align}
This angle is doping-dependent and near the SC optimal doping $(\delta \approx 1/8)$ this angle is negligible $\theta \ll \pi$. Starting from a lattice kinetic energy with up to 3rd nearest neighbor hoppings, we can effectively obtain the dispersion in the continuum limit. In the vicinity of region $1$ and $2$, this dispersion relation can be expanded up to quadratic order in the momentum deviation from the hotspots 
\begin{align}
    \epsilon_{1,\bk} = \epsilon_{2, -\bk} = v k_{x_1} + \gamma k_{y_1}^2, 
\end{align}
where the parameters $v$ and $\gamma$ label the hot-spots' Fermi velocity and band curvature, respectively. 
Due to the $C_4$ symmetry of the Fermi surface the dispersion relation around regions $3,4$ can be determined in terms of $v$ and $\gamma$ in a similar manner. We denote the UV momentum cutoff around the hot-spots by $\plambda \ll \pi a^{-1}$  and in order to be consistent in our approximations the curvature must satisfy $\gamma  \sim 1/\plambda$. 
For future convenience, we use the Fermi velocity and the lattice constant of copper-oxide planes $a$, to define an effective hopping amplitude, $t_h = v/a$, which forms a natural energy scale for the free electron dynamics. 
In what follows, we will take $\hbar = 1$. 

Above the SC phase transition temperature condensation occurs in the BDW channel such that the mean-field order parameter $\Phi_{\bQ}(\bk)$ acquires a non-zero expectation value. 
Fluctuations about this mean field solution describe the corresponding collective motions of the quasi-particles in this phase which can be coupled to external probes such as electromagnetic fields. 

As mentioned above, due to spin-exchange interactions, electrons in regions $1$ and $2$  are coupled to electrons in regions $3$ and $4$. In the BDW phase, this coupling can be described by the dynamical BDW pairing field $\Phi_{\bq}(\tau) = g\sum_{\bk}\psi_{4, \bk- \frac{\bq}{2}}^{\dagger}(\tau) \psi_{3, \bk + \frac{\bq}{2}}(\tau)$. 
Using this approximation the continuum limit of the t-J-V model in the BDW phase generates an interaction as follows
\begin{align}
    H^{\rho-\rho}_{\mathrm{int}} = -\sum_{\bk,\bk',\bq} g \Psi^{\dagger}_{\bk+\frac{\bq}{2}}\hat{V}\Psi_{\bk-\frac{\bq}{2}}\Psi^{\dagger}_{\bk'-\frac{\bq}{2}}\hat{V}\Psi_{\bk'+\frac{\bq}{2}}, 
    \label{density-density}
\end{align}
where the short-range interaction vertex $\hat{V}$ in terms of the Pauli matrices $\hsigma_{1,2,3}$ in the hot-spot basis is  $\hat{V} = \hsigma_{2}$. The associated BDW pairing field with this Hamiltonian reads 
\begin{align}
\Phi_{\bq}(\tau) = g\sum_{\bk}\psi_{2, \bk- \frac{\bq}{2}}^{\dagger}(\tau) \psi_{1, \bk + \frac{\bq}{2}}(\tau), 
\end{align}
where $\tau$ is the imaginary time. As a complex field $\Phi_{\bq}(\tau)$ can be decomposed into its Higgs \cite{pekker2015amplitude} and phason modes $(\Phi^{H}_{\bq}, -i\Phi^{G}_{\bq}) = (\Phi_{\bq} \pm \Phi_{-\bq}^{\dagger})/2$ which identify the amplitude and phase fluctuations, respectively. The latter is a current-carrying mode which is associated with the sliding motion of electrons \cite{lee1993conductivity, rice1979dynamics, browne1983collective}. 
Since these two kinds of fluctuations decouple from one another and only the latter is linearly optically active, in the following we only focus on phase fluctuations and henceforth take $\Phi_{\bq}$ to mean solely the phase-like component. 

Since our goal is the enhancement of SC in the regime where SC is the subleading instability, we consider temperatures above the SC transition temperature. 
Under these circumstances the hybridization of superconducting and BDW fluctuations can be ignored \cite{PhysRevB.92.184511}, and therefore, in the following we will suppress the spin indices.

\subsection{Phason Dispersion}

After condensation of electron-hole (e-h) pairs in the BDW phase, $\Phi_\mathbf{Q}(\mathbf{k}) = \phi P_\mathbf{Q}(\mathbf{k})$, with an amplitude $\phi$, the mean field Hamiltonian becomes 
\begin{equation}
H = \sum_{\bk}\Psi^{\dagger}_{\bk} \begin{pmatrix} 
\epsilon_{1\bk} &  \phi \\  \phi  & \epsilon_{2\bk}
\end{pmatrix} \Psi_{\bk} + \frac{1}{g}\phi^2.\label{MFHamil}
\end{equation}
Note that due to the d-wave symmetry of the ordering, e-h pairs around hot-spot regions $1,2$ and $3,4$ acquire the same condensation value $\phi$.
Hybridization of electrons in the bands $\epsilon_{1,2}$ results in an energy gap $d_{\bk}=\sqrt{(\epsilon^{d}_{\bk})^2 + \phi^2}$ where we have introduced the ``energy difference'' $\epsilon^{d}_{\bk} = (\epsilon_{2,\bk} - \epsilon_{1,\bk})/2$. After diagonalizing the above Hamiltonian the corresponding energy of the quasiparticles in the ``valence'' and ``conduction'' bands become 
\begin{equation}
    E^{c,v}_{\bk} = \epsilon^{m}_{\bk} \pm \sqrt{(\epsilon^{d}_{\bk})^2 + \phi^2},\label{QPDispersion}
\end{equation}
where the ``mean energy'' is defined as $\epsilon^{m}_{\bk} =(\epsilon_{2,\bk} + \epsilon_{1,\bk})/2$.

In order to study the dynamics of the BDW fluctuations, we need to go beyond the mean field Hamiltonian. This procedure may be done by applying a Hubbard-Stratonovich transformation to the interaction in Eq.\ref{density-density}. The resulting interaction is
\begin{align}
H_{\mathrm{e-\phi}} = \sum_{\bk,\bq}\Psi^{\dagger}_{\bk+\frac{\bq}{2}}\hh_{\bk,\bq}^{\mathrm{e-\phi}}\Psi_{\bk - \frac{\bq}{2} }+ \sum_{\bq}\frac{1}{g} \Phi_{\bq}^{\dagger}\Phi_{\bq}, \label{HS-Hamil}
\end{align}
where $\hh_{\bk,\bq}^{\mathrm{e-\phi}}=\hsigma_2 \Phi_{\bq}$.
To derive the dispersion relation of phason modes we calculate the two-particle Green's function in the BDW phase channel. In the usual way, the poles of this object correspond to the dispersion of the associated excitation, i.e. the BDW phase mode.
In the random phase approximation (RPA), this propagator is calculated by tracing out electrons from the effective action. The result can be compactly written $S^{\mathrm{eff}}[\Phi] = \sum_{q}D_q^{-1}\Phi_q\Phi_{-q}$ where we have employed combined notation for energy-momentum $q = (\omega, \bq)$, with bold symbols indicating the spatial momenta. As shown in the Methods and Materials section the BDW propagator $D_q$ receives contributions from both inter- and intra- band processes. However, in the low temperature limit, $T\ll\phi$, the latter are negligible and the leading order in frequency and momentum terms of the inverse propagator can be derived analytically 
\begin{align}
    D_q^{-1} &= \frac{1}{4 d_{\Lambda}^3}\left[v_{B}^{2}\left(q_x^2 + q_y^2 + 2q_xq_y\sin(2\theta_{\delta})\right) - \omega^2 \right] \notag \\
    & +  \mathcal{O}(\omega^4, \bq^4),
\end{align}
where $\thetadelta$ is the doping dependent in Eq.~\eqref{theta} and the BDW velocity is 
\begin{align}
     v_B = \frac{v}{\sqrt{2}}\left(1 + \frac{\phi^2}{\left(\phi^2 + v^2 \plambda^2 \right)} - \frac{4}{3v^4}\gamma^2 \plambda^2  \right)^{1/2},
\end{align}
which for small condensation fields asymptotically approaches $ v_B \approx v/{\sqrt{2}}$.
The $\sqrt{2}$ factor in $v_B$ is due to the fact that the total polarization bubble $\Pi_{q}^{\phi}$ contains contributions from both the regions $1,2$ and $3,4$.  
For doping levels in the range $\theta_{\delta} \in [0,\pi/4]$, the momentum dependence in the dispersion relation interpolates between the isotropic form $\left(q_x^2 + q_y^2\right)^{1/2}$ and $|q_x \pm q_y|$. 
The latter case is associated with vanishing values of doping where the Fermi surface becomes a $\pi/4$-rotated square in the Brillouin zone. In this limit the response function of BDW modes can be obtained without applying the hot-spot approximation (details can be found in the Supplementary Materials section). However, the former limit is most similar to the optimal doping regime in cuprates and will therefore be limit of main interest in this paper. 

\begin{figure}
\includegraphics[height=6cm]{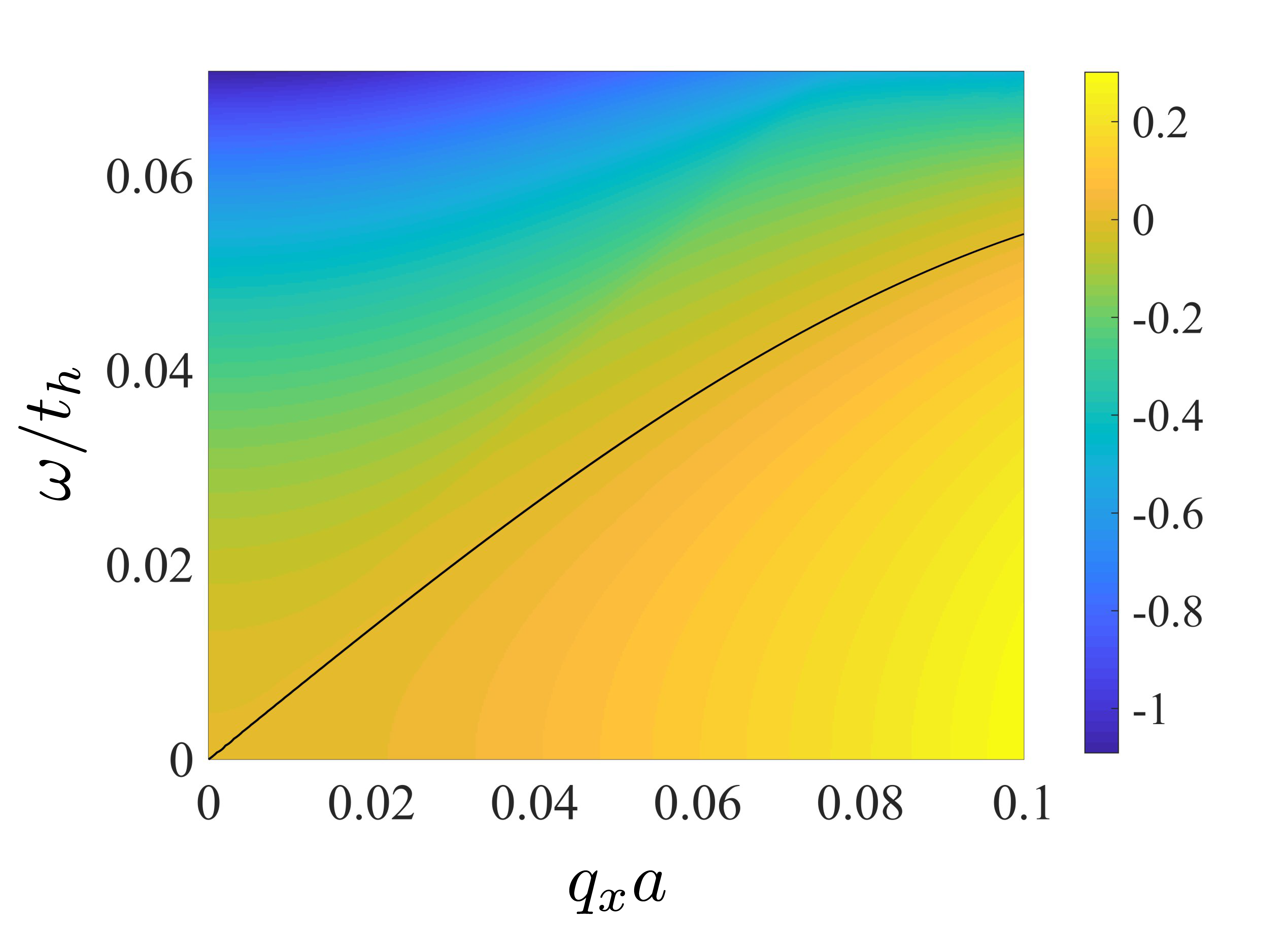}
\caption{
Intensity map of the inverse Green's function of bond density wave phason $D^{-1}(\omega, q_x)$ as a function of frequency and momentum $q_x$ for $q_y =0$. The solid line denotes the zeros of the inverse Green's function, corresponding to the dispersion of the phason mode. For this plot, we have chosen $\gamma[av] = 0.5, \phi[t_h] = 0.05, k_BT [t_h]=0.01$, where $a, v$ and $t_h$ label the lattice constant, Fermi velocity and the effective hopping amplitude, respectively. 
}\label{dispersionRel}
\end{figure}

In Fig.~\ref{dispersionRel} the dispersion relation which corresponds to the poles of $\mathrm{Re}(D_{q}^{\phi})$ is plotted as a solid black line.
As can be seen in this figure the linear dispersion relation is satisfied in the frequency regime $\omega < 2\phi$ corresponding to the regime where the frequency is within the BDW energy gap. 

\subsection{Phason-Photon Coupling}
The optical response of phason modes arises from the paramagnetic coupling of electronic current to the gauge potential $\bA$. From the standard minimal gauge coupling one finds the electron-photon interaction 
\begin{equation}
\hat{H}^{\mathrm{e-A}} = \sum_{\bk,\bq}\Psi^{\dagger}_{\bk+\frac{\bq}{2}}e \bA_{\bq}.\left(\bv_{m, \bk}\hat{I} - \bv_{d, \bk}\hat{\sigma}_3 \right) \Psi_{\bk - \frac{\bq}{2} }
,\label{Hamilt_e-A}
\end{equation}
where we have used the notation $\bv^{m,d}_{\bk} = \nabla_{\bk}(\epsilon^{m}_{\bk}, \epsilon^{d}_{\bk})$ for the mean and difference Fermi velocities, respectively.
This leads to a net paramagnetic response for phason modes which can be expressed as
\begin{align}
    S_{\bA-\phi} = \sum_{q} ie\bLambda_{q}.\left(\bA_{-q}\Phi_{q} - \bA_{q}\Phi_{-q}\right)
\end{align}
where $\bLambda_{q} = \bLambda_{q}^{m} + \bLambda_{q}^{d}$ is composed of the ``mean'' and ``difference'' momentum-dependent optical coupling strengths of the phason field, which receive contributions from $\bv^{m}_{\bk}, \bv^{d}_{\bk}$ velocity vertices, respectively.
At low temperatures, $T\ll\phi$, and for negligible $\thetadelta$ the leading order terms in the gradient expansion of the optical couplings take the form
\begin{gather}
    \bLambda^{d}_{\bq} =  - \frac{2 v \omega}{\phi d_{\Lambda}}(\hat{\bx} - \hat{\by})\label{interactionVertex} \\  
    \bLambda^{m}_{\bq} =  \frac{4 \gamma^2 v \plambda^2 \left(3 \phi^2 + 2 v^2 \plambda^2\right)}{9 \phi^3 d_{\Lambda}^3} \omega q_x q_y (\hat{\bx} - \hat{\by}),
\label{opticalVertex}
\end{gather}
where $d_{\Lambda} = \sqrt{\phi^2 + v^2 \plambda^2}$. Note that the interaction vertex vanishes in the limit of DC fields and its dominant component is parallel to the BDW wavevector $\bQ$. Also, from the momentum dependence of the ``mean'' term and its relative sign with respect to the ``difference'' term it can be deduced that at a fixed wavevector $|\bq|$ the associated effective coupling strength is strongest when $q_x = -q_y$. 

\subsection{Optical Conductivity}
While the ``local'' non-equilibrium optical conductivity of superconductors has been recently studied \cite{kennes2017nonequilibrium}, here we are primarily interested in the ``non-local'' behavior of the conductivity associated with excitation of the BDW collective modes when the resonance condition is satisfied. 
Having calculated the coupling interaction vertex and the self energy of phason modes the non-local conductivity can be extracted in a straightforward manner by integrating out the BDW fields and obtaining an effective action for photons. The Feynman diagrams required in the calculation of the optical response of collective modes \cite{lee1993conductivity}, are presented in Fig.~\ref{FeynmanDiagram}. 
The final results of these diagrams for the complex conductivity is 
\begin{align}
    \sigma_{\alpha \beta}(\omega, \bq)  = \frac{i e^2 \Lambda_{q}^{\alpha} \Lambda_{-q}^{\beta} D_q}{\omega}, 
\end{align}
where $\alpha, \beta = \{x, y\}$. Inserting the results of Eqs.~\eqref{interactionVertex} and~\eqref{opticalVertex} in the conductivity tensor, we conclude that at low-temperature the conductivity tensor satisfies $\sigma_{xx}= \sigma_{yy} = -\sigma_{xy} \equiv \sigma_{\mathrm{diag}}$. Hence, from Ohm's law, $J_{\alpha} = \sigma_{\alpha \beta} E_{\beta}$, $\mathbf{J}$ and $\mathbf{E}$ being the current density and the electric field, respectively, it can be inferred that the maximum optical dissipation can be obtained when the electric field of the laser is along the diagonal or more generally parallel to the BDW wavevector $\bQ$. 

\begin{figure}
\includegraphics[height=5.8cm]{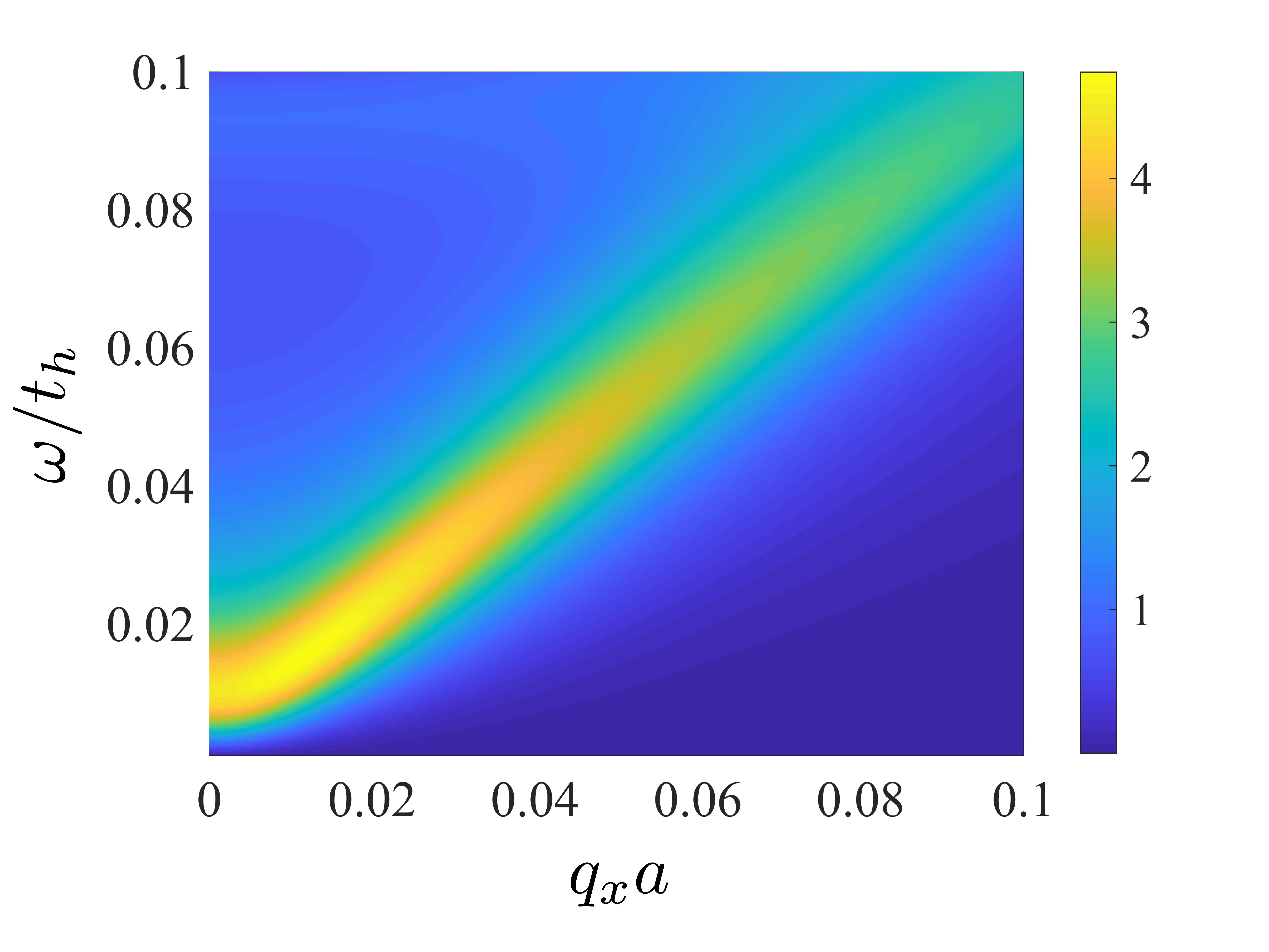}
\caption{
Intensity map of the resistive part of the conductivity $\operatorname{Re}\sigma_{\mathrm{xx}}(\omega, \bq)=\sigma_{\mathrm{yy}}(\omega, \bq)$, for diagonal momentum $q_x = q_y$ and model parameters $ \gamma[av] = 0.5, \phi[t_h] = 0.05, k_B T[t_h]=0.01$.
The peak response, indicating maximum power absorption, indicates resonance with the BDW phason mode.
The appearance of gap in the conductivity at low momenta, even though the phason is still gapless, is due to the vanishing of phason-photon coupling for small $q$, cf. Eq.~\eqref{interactionVertex}. }\label{opticalCond}
\end{figure}

In Fig.~\ref{opticalCond}, we have plotted the real part of the diagonal optical conductivity $\sigma_{\mathrm{diag}}(\omega, \bq)$, which determines the dissipated power in the system, according to the Joule's law, at a non-zero temperature. 
From this plot, we can determine the optimal driving frequency for melting of the BDW modes. 
Note that in this figure the low-frequency and low-momentum behavior of the optical conductivity qualitatively differs from the behavior observed in Fig.~\ref{dispersionRel}. This is due to the additional dependence of the conductivity on the optical interaction vertices and their non-trivial frequency-dependence, which was derived in Eqs.~\eqref{interactionVertex} and~\eqref{opticalVertex}. In general, there are additional paramagnetic and diamagnetic contributions, which are insignificant under the resonance condition, and therefore, ignored in our study.

Besides numerical computations the conductivity can be analytically evaluated for low frequencies at zero temperature 
\begin{align}
    \sigma_{\mathrm{diag}}(\omega, \bq) = \frac{16 i e^2 v^2 \omega d_{\Lambda}}{\phi^2 \left[v_B^2 q^2 - (\omega + i \eta)^2\right] }, \label{cond}
\end{align}
where $q^2 = q_x^2 + q_y^2$ for vanishing $\thetadelta$. Note that we have retained a positive infinitesimal imaginary shift $\eta$ in the frequency, to explicitly demonstrate the retarded analytic structure of this result. 

\subsection{BDW Melting Rate}

Once the real part of conductivity is inserted into Joule's law, the absorbed power density can be evaluated. In the zero-temperature limit where the optical conductivity is given by Eq.~\eqref{cond} and for a resonant narrow-linewidth laser field with a vector potential amplitude of $A_0$ and frequency $\omega_0$ the time-averaged absorbed power density can be expressed as
\begin{align}
    \bar{P} = A_0^2 \omega_0^2 \frac{4 \pi v^2  d_{\Lambda}}{\phi^2 }\delta(\omega_0- v_B q), \label{power}
\end{align}
In this relation the Dirac $\delta$-function encodes the spectral behavior of the BDW collective modes and appears due to the infinitessimal quantity $\eta \rightarrow 0^+$ in Eq.~\eqref{cond}. Note that once the resonance condition for the laser wavelength is satisfied the delta function can be integrated over which results in a finite value for the dissipated power density.

When the resonance condition is satisfied the energy absorbed from the drive allows for electrons to leave the condensed BDW state.
One can obtain a phenomenological upper bound for the melting rate of the BDW order parameter by assuming that after leaving the BDW phase electrons do not return to this state, i.e. equating the optical power density with the rate of change in the mean-field energy density of the BDW phase
\begin{align}
    \frac{d}{dt}\Big(\frac{\phi^2}{g}\Big) = -\frac{1}{2}A_0^2 \omega^2 \mathrm{Re}\{\sigma(\omega, \bq)\}.\label{BDWMelting}
\end{align}
Provided that the initial value of the order parameter is sufficiently large, a low-temperature approximation for the optical conductivity is permissible in the early stage of the melting process.
In this regime Eq.~\eqref{BDWMelting} can be integrated  
\begin{align}
    \phi(t) = \left(\phi_0^4 - rt\right)^{1/4},
\end{align}
where we have defined $r\equiv 8 \pi  g \dlambda v^2 A_0^2 \omega_0^2$. For a small irradiation power this relation can be further recast to a linear decay with a rate equal to $r/4\phi_0^3$. In SI units, for a laser field with a frequency around $5$ THz and an electric field of $10^7$ V/m, the melting time scale $t_{\mathrm{melt}} \simeq \phi_0^4/r$ is in the range of picoseconds, comparable to what has been observed for the melting time scale observed when melting via phonon excitation \cite{fausti2011light,Forst2014}. 

\begin{figure}
\includegraphics[height=6cm,keepaspectratio]{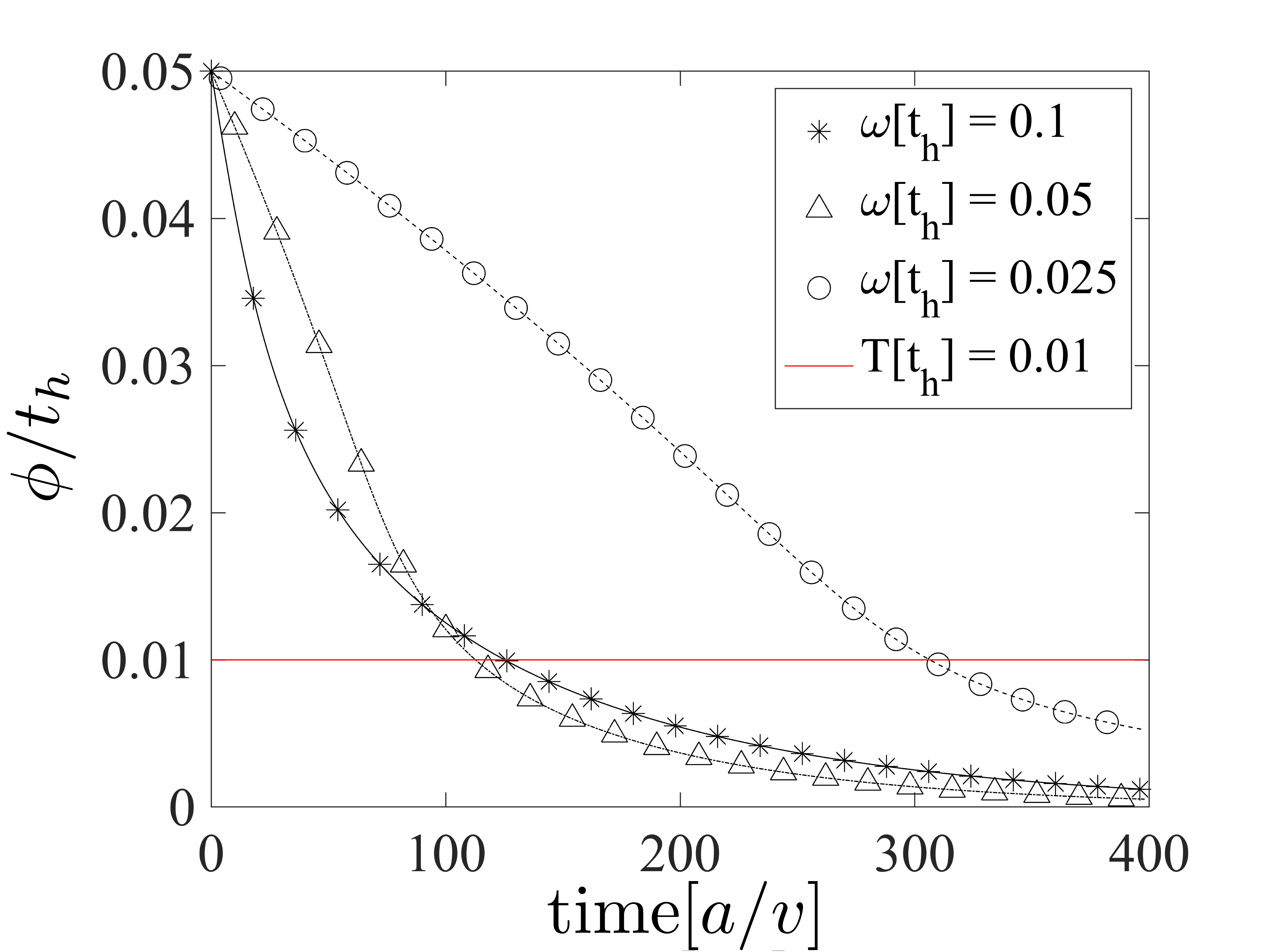}
\caption{Melting of the BDW order parameter as a function of time.
The time evolution is plotted for three different frequencies of the laser field with the photon momentum being determined by the resonance condition. The red line shows the equilibrium temperature at which the melting process is carried out, for reference.}
\label{PhiOfT}
\end{figure}

To investigate the time dependence of the order parameter beyond this limit we numerically solve Eq.~\eqref{BDWMelting}. Fig.~\ref{PhiOfT} displays the results of this computation when the laser field's frequency and wavevector are resonant with the phason absorption and the initial value of the order parameter is chosen such that we are deep inside the BDW phase. Moreover, we utilize the fact that optical coupling can be maximized for diagonal wavevectors $q_x = \pm q_y$. In this figure three curves are displayed, corresponding to different laser frequencies while other parameters are kept equal. As explained above for early times the melting process demonstrates a nearly linear decay. This behavior is followed by a saturating behavior when the order parameter becomes comparable to the temperature. In this regime the non-trivial $\phi$-dependence of the optical conductivity starts to emerge which results in an overall deceleration of the melting process. Also, comparison of different curves indicates that since the power density increases with the frequency the order parameter decay rate is higher at larger frequencies. 

It should be borne in mind that in any experimental realization of this technique a finite fraction of the electrons after being driven out from the BDW phase will eventually re-condense into the BDW phase due to the subsequent electron-electron and electron-phonon interactions. Hence, our phenomenological computation is an upper bound on determining the melting rate of the BDW order parameter.  

\subsection{Plasmonic\ Engineering}
Since the BDW phase transition temperatures could reach around $150K$, the proposed frequency regime in our approach is $\omega \lesssim 5 $THz. In this frequency regime doped graphene hosts plasmons that simultaneously have low losses and significant wavelength confinement ratios.
For a graphene sheet mounted in a substrate with a relative dielectric constant $\varepsilon_r$, at a given frequency $\omega$ the SPP confinement can be approximated as
\begin{align}
\frac{\lambda_{\mathrm{sp}}}{\lambda_o} \approx \alpha \frac{4 \mu}{\varepsilon_r} \frac{1}{\left(\omega + i \tau^{-1}\right)}\label{plasmonFreq}
\end{align}
where the fine structure constant is $\alpha = e^2/4\pi \varepsilon_0  \approx 1/137$, and $\tau$ and $\mu$ denote the electron relaxation time and the chemical potential of doped graphene, respectively \cite{jablan2009plasmonics, luo2013plasmons}.
This relation is a reasonable approximation provided the frequency of photons is smaller than the optical phonon frequency of graphene $\omega_{Oph}\approx 50$THz which is compatible with our proposal. 

Eq.~\eqref{plasmonFreq} indicates that the wavelength of plasmons can be adjusted by varying the doping level of graphene or using different dielectric substrates. Recently, it has been observed that this confining ratio can be amplified as high as 300 \cite{lundeberg2017tuning} for frequencies as low as a few THz which is close to the confinement potential required for our proposal. An even higher confinement ratio of up to nearly $10^3$ is reported in other van der Waals materials \cite{basov2016polaritons, iranzo2018probing}. 
In the relevant frequency regime for our system, topological materials such as $\mathrm{Bi}_2\mathrm{Te}_3$ accommodate surface plasmon-polariton modes, albeit with a relatively low quality factor. Nevertheless, since we can consider an array of laser fields, this drawback does not affect our experimental setup. 
Therefore, the frequency and momentum matching in our desired frequency regime is accessible with the current technology.

\section{Discussion}
We have presented here a scheme for enhancing superconductivity via the resonant melting of a competing order.
While in this work we have not directly considered the interplay between irradiation and superconductivity, it should only play a secondary effect on the applicability of our proposal. 
This is because, in contrast to the BDW order, which couples linearly to light, superconductivity can only couple quadratically to photons due to gauge invariance of the electromagnetic field. Consequently, such processes are beyond linear response and the optical power that could be dissipated in the superconducting condensate is quartic in the amplitude of the laser field, in contrast to the quadratic behavior we derived in Eq.~\eqref{power}. 
Furthermore, by tuning the energy and momentum of light to be resonant with BDW phason modes, we expect the response of the BDW to be much stronger than the non-resonant processes which are unfavorable for superconductivity.
By controlling such adverse effects, via reservoir engineering techniques, we expect that the net effect of our resonant irradiation protocol would be an enhancement of superconductivity.

Finally, we point out that the same technique may be applied not only to cuprate materials, but also to other strongly correlated systems such as iron-pnictide superconductors or bismuthates \cite{nicoletti2017anomalous} where two or more competing or intertwined orders are coexistent. More generally, such optical pumping schemes could become a new tool in manipulation of correlated states of electrons. 

\section*{Methods}

To diagonalize the original mean field Hamiltonian in Eq.\eqref{MFHamil}, we apply a unitary transformation, $U_{\bk}$, to the original spinor. The rotated spinor is denoted by $\tilde{\Psi}_{\bk}  = U_{\bk} \Psi_{\bk}$. The rotation matrix up to gauge choice is
\begin{align}
U_{\bk} = \begin{pmatrix}
\nu_{\bk} & -u_{\bk} \\ u_{\bk} & \nu_{\bk}
\end{pmatrix}, 
\end{align}
where in the above the rotation coefficients are
\begin{align}
u_{\bk} = \frac{r_{\bk}}{(1 + r_{\bk}^2)^{1/2}}, \quad \nu_{\bk} = \frac{1}{(1 + r_{\bk}^2)^{1/2}}, 
\end{align}
and $r_{\bk} = \phi/(\epsilon_{d,\bk} + d_{\bk})$. Next, the combined interaction Hamiltonian in Eq.~\eqref{HS-Hamil} and \eqref{Hamilt_e-A} is transformed to the new basis 
\begin{align}
    H^{\mathrm{int}} = \sum_{\bk,\bq}\tilde{\Psi}^{\dagger}_{\bk+\frac{\bq}{2}}\tilde{h}_{\bk,\bq}^{\mathrm{int}}\tilde{\Psi}_{\bk - \frac{\bq}{2} },
\end{align}
where 
\begin{align}
    \tilde{h}^{\mathrm{int}} & = \Big(e\bA_{\bq}.\bv_{m, \bk} w^{(0)}_{\bk,\bq} +i w^{(2)}_{\bk,\bq}\Phi_{\bq}^G \Big)\ \hsigma_0  - e w^{(1)}_{\bk,\bq} \bA_{\bq}.\bv_{d, \bk} \hsigma_1  \notag \\ 
& + \Big(ie\bA_{\bq}.\bv_{m, \bk}w^{(2)}_{\bk,\bq} + w^{(0)}_{\bk,\bq}  \Phi_{\bq}^G \Big)\ \hsigma_2 - e w^{(3)}_{\bk,\bq}\bA_{\bq}.\bv_{d,\bk}\hsigma_3
\end{align}
and in the above equation we have defined the coherence factors $w_{\bk,\bq}^{(i)}$ as
\begin{subequations}
\begin{align}
&w^{(0)}_{\bk,\bq} = \nu(\bk^+)\nu(\bk^-) + u(\bk^+)u(\bk^-), \\ &w^{(1)}_{\bk,\bq} = u(\bk^+)\nu(\bk^-)+\nu(\bk^+)u(\bk^-), \\
& w^{(2)}_{\bk,\bq} = \nu(\bk^+)u(\bk^-) - u(\bk^+)\nu(\bk^-), \\  &w^{(3)}_{\bk,\bq} =\nu(\bk^+)\nu(\bk^-)- u(\bk^+)u(\bk^-),
\end{align}
\end{subequations}
where $\bk^{\pm} = \bk \pm \bq/2$. 

\begin{figure}
\includegraphics[height=4cm,keepaspectratio]{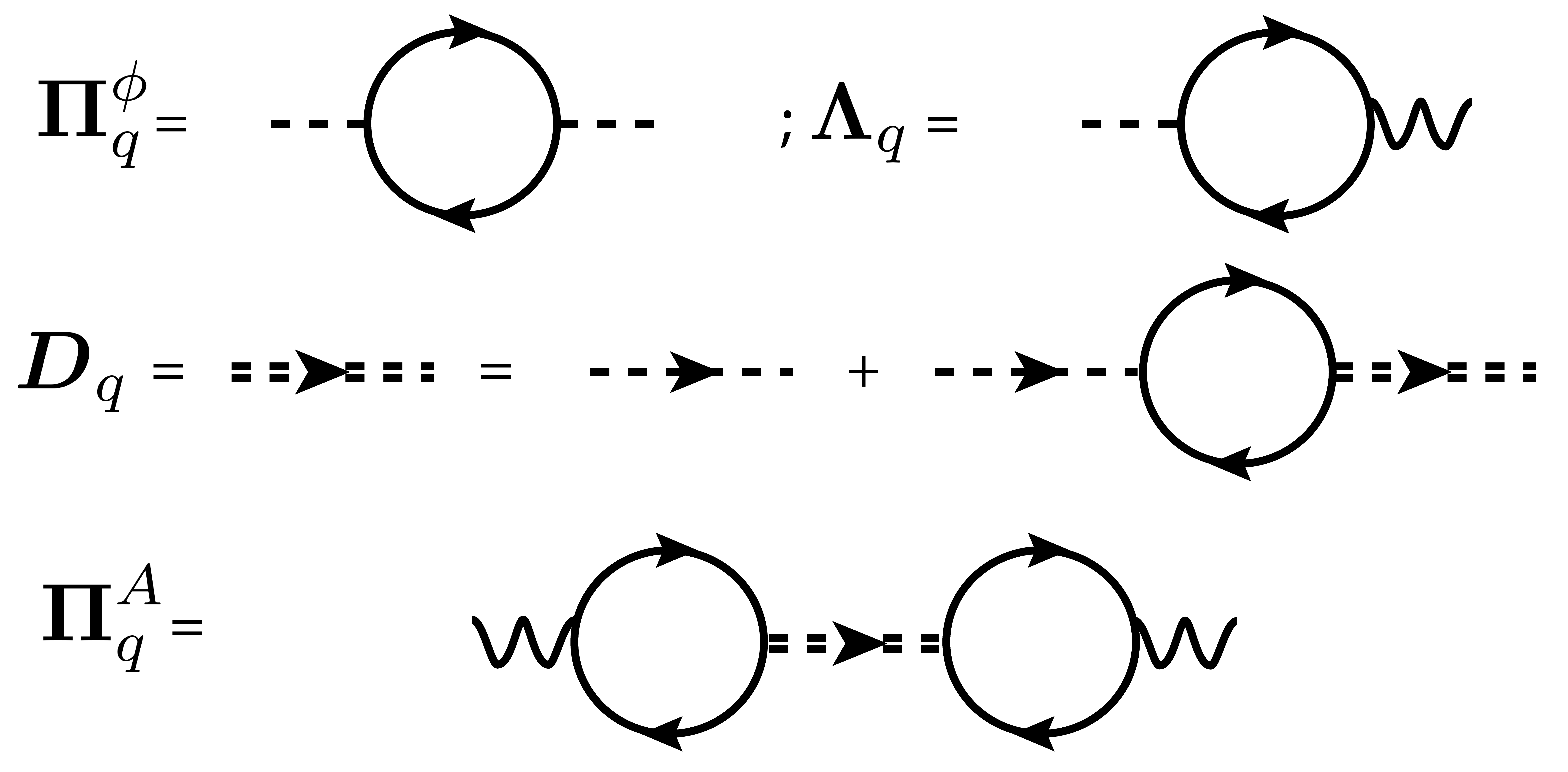}
\caption{
Feynman diagrams of polarization functions. Solid, dashed and wiggly lines represent electrons', phasons' and photons' propagators, respectively. Top row: On the left the polarization diagram of phason modes, and on the right the phason-Photon interaction vertex. Middle row: phason propagator obtained from a Dyson equation summation of phason polarization diagrams. 
Bottom row: Collective mode's contribution to the conductivity by exchanging phason fields. 
}\label{FeynmanDiagram}
\end{figure}

The building blocks of the required Feynman diagrams in Fig.\ref{FeynmanDiagram}, are the e-h correlation functions. These correlations can be calculated in the imaginary time Matsubara formalism.
We introduce Matsubara frequencies which at inverse temperature $\beta =1/T$ are given by $\omega_m = 2\pi m T$ and $\epsilon_n = 2\pi (n+1/2) T$ for bosonic and fermionic fields, respectively.
Ultimately, in order to obtain the causal (retarded) response of the system, we will analytically continue to real frequencies $i\omega_m \rightarrow \omega + i \delta^{+}$. Introducing the fermionic Green's functions $G_{n,\bk} = (i\epsilon_n- \epsilon_a(\bk))^{-1}$ the e-h correlations are obtained after the frequency summation   
\begin{align}
\chi_{q,\bk}^{a, b} & = T \sum_{n}G_{n+m,\bk + \frac{\bq}{2}}G_{n,\bk - \frac{\bq}{2}} \notag \\ 
& =\frac{n_{F}(E^b_{\bk+\frac{\bq}{2}}) - n_{F}(E^a_{\bk-\frac{\bq}{2}})}{i\omega_m + E_{\bk+\frac{\bq}{2}}^{b} - E_{\bk-\frac{\bq}{2}}^{a} },
\end{align}
where the Fermi-Dirac distribution $n_F(E) = (1 + e^{\beta E})^{-1}$ is obtained after performing the fermionic Matsubara frequency summation and $a,b = \{v,c\}$ indexes the lower (valence) and upper (conduction) bands' eigenenergies according to Eq.~\eqref{QPDispersion}.

From here, one can find the saddle point of the partition function which requires the mean field BDW order parameter to satisfy 
\begin{equation*}
    \frac{1}{g}=  \sum_{\bk}\frac{1}{2d_{\bk}} \Big( n_F(E_{\bk}^v) - n_F(E_{\bk}^c) \Big), 
\end{equation*}
where the momentum summation is limited to the UV momentum cutoff around the hot-spots $\sum_{\bk} = \int d^2 \bk/ (2\plambda)^2$.

To take into account the collective modes, we must include Gaussian fluctuations of electrons above the mean field solution.
To do so, we integrate out the electrons in the partition function, which gives rise to a number of terms in the effective action for the phason modes and their interaction with photons.
Decomposing the total interaction Hamiltonian in terms of the Pauli matrices $\tilde{h}^{int} = \tilde{h}^{int}_{\alpha}\hsigma_{\alpha}$, and using the energy-momentum convention $q = (\omega_m, \bq)$, the effective action is calculated in the rotated basis in terms of the Green's functions 
\begin{align}
    S^{\eff}_{\alpha \beta} = \frac{1}{2}\sum_{k,q}\tr\left(\tilde{G}_{k+\frac{q}{2}}\tilde{h}^{\mathrm{int}}_{q; \alpha }\hsigma_{\alpha} \tilde{G}_{k-\frac{q}{2}} \tilde{h}^{\mathrm{int}}_{-q; \beta}\hsigma_{\beta} \right),\label{effectiveAction}
\end{align}
where the trace is taken over fermionic degrees of freedom. It should be emphasized that in the above formalism all the different combinations of Pauli matrix indices $\alpha$ and $\beta$ which contribute to a single physical process must be included. This functional approach is equivalent to employing the RPA to evaluate the resulting Green's function of phason fields which is plotted on the top line of Fig.~\ref{FeynmanDiagram}. Considering the coupling constant $g$ as the non-interacting Green's function of the phase mode $D^0$, the Dyson equation is
\begin{align}
    D_{q}^{-1} &= (D^{0}_q)^{-1} + \Pi^{\phi}_{q} \notag \\
    &=  \frac{1}{g} + \sum_{\bk} \left[ \frac{1}{4} \left( (1 - \tilde{f}_{\bk, \bq}^{(\phi)} )(\chi^{vv}_{q} + \chi^{cc}_{q}) 
 \notag \right. \right.\\ 
 & \left. \left. + (1 + \tilde{f}_{\bk, \bq}^{(\phi)} ) (\chi^{vc}_{q} + \chi^{cv}_{q}) \right)\right],
\end{align}
where $\tilde{f}_{\bk, \bq}^{(\phi)}$ are coherence factors originating from the rotation of spinors from the $1,2$ basis to the $v,c$ bands,
\begin{align}
\tilde{f}_{\bk,\bq}^{(\phi)} = \frac{ \phi^2 + \epsilon_{\bk + \frac{\bq}{2}}^{d} \epsilon_{\bk - \frac{\bq}{2}}^{d}}{ d_{\bk + \frac{\bq}{2}}d_{\bk - \frac{\bq}{2}} }.
\end{align}

Next, the optical coupling of BDW fields can be expressed in terms of the e-h correlations
\begin{align}
    \Lambda^{m}_{\bq}  = & \sum_{\bk} \bv^{m}_{\bk}\frac{\phi   (\epsilon_{\bk + \frac{\bq}{2}}^{d}    -   \epsilon_{\bk - \frac{\bq}{2}}^{d})}{2 d_{\bk + \frac{\bq}{2}}   d_{\bk - \frac{\bq}{2}}}  \left(\chi_{\bk, q}^{vv} + \chi_{\bk, q}^{cc} - \chi_{\bk, q}^{vc} - \chi_{\bk, q}^{cv}\right)
\end{align}
and 
\begin{align}
    \Lambda^{d}_{\bq}  = & 
    \sum_{\bk} \bv^{d}_{\bk} \left[\left( \frac{\phi}{2 d_{\bk -  \frac{\bq}{2}}} - \frac{\phi}{2 d_{\bk + \frac{\bq}{2}}} \right)  (\chi_{\bk, q}^{vv} - \chi_{\bk, q}^{cc}) \notag \right.\\
    & \left. - \left( \frac{\phi}{2 d_{\bk+ \frac{\bq}{2}}}  +  \frac{\phi}{2 d_{\bk - \frac{\bq}{2}}} \right) \left(\chi_{\bk, q}^{vc}  - \chi_{\bk, q}^{cv}\right) \right].
\end{align}
As it is illustrated in the bottom line of Fig.~\ref{FeynmanDiagram} this interaction vertex appears in photons' polarization tensor. Notice that the related process only includes the collective contributions to the susceptibility originating from the exchange of BDW phason modes.
\begin{align}
    \Pi_{\alpha \beta}^{A}(q) &= \frac{\delta^2 S_\mathrm{eff}[A]}{\delta A^{\alpha}_q \delta A^{\beta}_{-q}} \notag \\
    & = e^2 \Lambda^{\alpha}_q \Lambda^{\beta}_{-q} D_q,
\end{align}
where $\alpha, \beta =\{x, y\}$. The associated non-local conductivity is given by $\sigma_{\alpha \beta}(\omega,\bq) = i\Pi_{\alpha \beta}^A(\bq, \omega)/\omega$. 
For $\thetadelta\approx 0$ only the diagonal terms of this tensor have a significant value.

For the numerical computation of the retarded Green's functions in the frequency domain we apply a shift along the vertical axis $\omega \rightarrow \omega + i\eta $. Numerically, this leads to a Lorentzian approximation for the real part of the conductivity. 

\paragraph{Energy Absorption Rate.---}
For a general probe field $\hat{Y}$ we associate a coupling Hamiltonian. We define the susceptibility of operator $\hat{X}$ with respect to the operator $\hat{A}$ as 
\begin{align}
\chi_{XY}(t,t') = i \theta(t -t') \langle [\hat{X}(t), \hat{Y}(t')] \rangle
\end{align}
For a sinusoidal perturbation $f(t) =f_0 \cos(\omega t)$ at frequency $\omega_0$ and with amplitude $f_0$, the calculation of the energy absorption rate yields
\begin{align}
\overline{Q} &= \frac{1}{2}f_0^2 \omega_0\ \mathrm{Im}\{\tilde{\chi}_{XY}(\omega_0)\}. 
\end{align}
where $\tilde{\chi}_{XY}$ denotes the Fourier transform of the susceptibility.
The appropriate susceptibility in this problem is the polarization tensor $\Pi^{\bA}_q$. Next, by equating the energy absorption rate with the melting rate of the ground state energy we obtain a non-linear differential equation for the time dependence of the BDW field $\phi(\tau)$ which we solve by the Runge-Kutta method.

\begin{acknowledgments}
This research was supported by  supported by US-ARO (contract No. W911NF1310172) (Z.R.), NSF DMR-1613029, DARPA DRINQS project FP-017, ``Long-term High Temperature Coherence in Driven Superconductors,'' and Simons Foundation (V.G.), and AFOSR FA9550-16-1-0323, ARO W911NF-15-1-0397, and NSF Physics Frontier Center at the Joint Quantum Institute (H.D. \& M.H.). 

\end{acknowledgments}
\bibliography{mybib.bib}

\setcounter{equation}{0}
\renewcommand{\theequation}{A.\arabic{equation}}

\begin{widetext}

\section{Appendix}

\subsection{Phason dispersion near half-filling}
In this section, we present our results for the dispersion relation of BDW phasons near half filling. In this limit, we are able to go beyond a low energy ``hot-spot'' model.
However, since the dispersion relation of the quasiparticles has nodal lines in the Brillouin zone the momentum integrals appearing in the particle-hole correlations need to be regularized. 
At low dopings the Hubbard model on a square lattice is susceptible to the formation of d-wave charge density waves at wavevector $\bQ=(\pi, \pi)$ \cite{chakravarty2001hidden}.
Therefore, in the following, we assume a density-density interaction whose dominant mean-field solution is a d-wave BDW state 
\begin{align}
    H^{\rho-\rho}_{\mathrm{int}} = -\sum_{\bk,\bk',\bq} g f_{\bk} f_{\bk'}  c^{\dagger}_{\bk-\frac{\bQ-\bq}{2};\sigma}c_{\bk+\frac{\bQ-\bq}{2};\sigma}c^{\dagger}_{\bk'+\frac{\bQ-\bq}{2};\sigma'}c_{\bk'-\frac{\bQ-\bq}{2};\sigma'}, 
\end{align}
where $f_{\bk}$ is the d-wave symmetry factor $f_{\bk} = \cos(k_x)-\cos(k_y)$ and $\bQ=(\pi, \pi)$. It should be borne in mind that unlike the hot-spot model, here, the momentum summations are initially evaluated over the whole Brillouin zone.
However, BDW phase, in order to avoid overcounting the fermionic degrees of freedom, the new Brillouin zone should be identified with the magnetic Brillouin zone of the original square lattice model to reflect the period doubling due to symmetry breaking.
Doing so introduces an additional index for fermions, describing A or B site within the period-doubled lattice.
The BDW order parameter is then defined as 
\begin{align}
    \Phi_{\bq} = g\sum_{\bk} f_{\bk} c^{\dagger}_{\bk+\frac{\bQ-\bq}{2}}c_{\bk-\frac{\bQ-\bq}{2}},
\end{align}
where we sum over all internal fermionic indices.
After condensing the interaction Hamiltonian at $\langle \Phi_{\bq=0}\rangle = \phi$ the mean-field Hamiltonian becomes
\begin{align}
H_{\mf} = \sum_{\bk} \Psi_{\bk}^{\dagger} \begin{pmatrix}
\epsilon_{\bk-\frac{\bQ}{2}} & -f_{\bk} \phi \\
-f_{\bk} \phi & \epsilon_{\bk+\frac{\bQ}{2}}
\end{pmatrix}
\Psi_{\bk} + \frac{1}{g}\phi^2, 
\end{align}
where $\Psi_{\bk}^{\dagger} = (c_{\bk-\frac{\bQ}{2}}^{\dagger}, c_{\bk+\frac{\bQ}{2}}^{\dagger})$. Also, the dispersion relation can be described by the first and second nearest neighbor hoppings $t$ and $t'$,
\begin{align}
\epsilon_{\bk} = - 2t (\cos{k_x a} +\cos{k_y a})+ 4t'\cos(k_x a)\cos(k_y a).
\end{align}
Notice that the only difference of this MF Hamiltonian and the hot-spot model is the fact that the symmetry factor $f_{\bk}$ appears in the off-diagonal component of the Hamiltonian vertex. The energy of the quasiparticles when the nearest neighbor hopping is included reads
\begin{align}
    d_{\bk} = \sqrt{(\epsilon^{d}_{\bk})^2 + f_{\bk}^2 \phi^2}
\end{align}
where $\epsilon^{m,d}_{\bk} = \frac{1}{2}(\epsilon_{\bk+\bQ/2 } \pm \epsilon_{\bk-\bQ/2 })$.
The self-consistency equation for the coupling strength $g$ is obtained at the saddle point of the effective action 
\begin{align}
    g^{-1} =  \sum_{\bk}\frac{ f_{\bk}^2 }{2d_{\bk}} \Big( n_{F}(-d_{\bk}) - n_{F}(d_{\bk}) \Big) .
\end{align}
The BDW dispersion and its optical coupling can be studied by incorporating the interaction fluctuations over the MF solution 
\begin{align}
    H_{\mathrm{int}}= \sum_{\bk, \bq}\Psi_{\bk+\frac{\bq}{2}}^{\dagger}  \begin{pmatrix} e\bA_{\bq}.(\bv_{\bk}^{m} - \bv_{\bk}^{d}) &  -i f_{\bk}\Phi_{\bq} \\
i f_{\bk} \Phi_{\bq} & e\bA_{\bq}.(\bv_{\bk}^{m} + \bv_{\bk}^{d}) 
\end{pmatrix}\Psi_{\bk-\frac{\bq}{2}}. 
\end{align}
The calculation of the Green's function of the phason fields is similar to the main text's derivation and follows from  Eq.\ref{effectiveAction}
\begin{align}
    D_{q}^{-1} & = (D^{0}_q)^{-1} + \Pi^{\phi}_{q} \notag \\
    & =  \frac{1}{g} + \sum_{\bk} \frac{ f_{\bk}^2 }{4} \left[ \left(1 - \tilde{f}_{\bk, \bq}^{(\phi)} \right)\left(\chi^{vv}_{q} + \chi^{cc}_{q} \right) 
+ \left(1 + \tilde{f}_{\bk, \bq}^{(\phi)} \right) \left(\chi^{vc}_{q} + \chi^{cv}_{q} \right) \right].
\end{align}
As before the coherence factors $\tilde{f}_{\bk, \bq}^{(\phi)}$ display the rotation of spinors from the $1,2$ basis to the $v,c$ bands,
\begin{align}
\tilde{f}_{\bk,\bq}^{(\phi)} = \frac{ \phi^2 f_{\bk + \frac{\bq}{2}} f_{\bk - \frac{\bq}{2}}+ \epsilon_{\bk + \frac{\bq}{2}}^{d} \epsilon_{\bk - \frac{\bq}{2}}^{d}}{ d_{\bk + \frac{\bq}{2}}d_{\bk - \frac{\bq}{2}} }.
\end{align}
The phason's dispersion relation is determined from the poles of $D_{q}$. At the zero-temperature limit these poles can be calculated analytically by expanding the Green's function in powers of the momentum $\bq$. This expansion is more conveniently performed in a $\pi/4$-rotated basis $\tilde{q}_{x,y} = q_x\pm q_y$, because the momentum summation must be performed in the magnetic Brillouin zone $|\tilde{q}_{x,y}| \le  \pi$,
\begin{align}
   D_q^{-1} &= \sum_{\bk} \frac{ f_{\bk}^2 }{8 d_{\bk}^3}\left[- \omega^2 + \frac{d_{\bk}}{2}  \tilde{q}_{ij}\tilde{\partial}_{ij}d_{\bk} + \frac{3 \omega^2 \tilde{q}_{ij}\tilde{\partial}_{ij} d_{\bk}}{8 d_{\bk}} - \frac{1}{16} \left(\tilde{q}_{ij}\tilde{\partial}_{ij}d_{\bk}\right)^2 + \frac{d\tilde{q}_{ijmn} \tilde{\partial}_{ijmn}d_{\bk}}{96 }  \right] \\
   & +  \mathcal{O}\left(\omega^4, \bq^6 \right), 
\end{align}
where we have introduced the compact notation $\tilde{q}_{ij} = \tilde{q}_i\tilde{q}_j$, and $\tilde{\partial}_{ij} = \frac{\partial^2}{\partial \tilde{k}_i \partial \tilde{k}_j}$. Also according to the Einstein notation a summation over the repeated coordinate indices is presumed. 
Note that the momentum integrals in the above are divergent. This is because the energy gap has nodal lines along $k_x = - k_y$
\begin{align}
    d_{\bk} = \sqrt{ \phi^2 \left(\cos(k_x) - \cos(k_y)\right)^2 + t^2 \left(\sin(k_x) + \sin(k_y)\right)^2}
\end{align}
Furthermore, it can be shown that inclusion of longer-range hopping amplitudes when constrained to preserve the $C_4$ symmetry of the square lattice does not eliminate this nodal line. Therefore, to regularize the integrals we consider a small $C_4$-symmetry breaking hopping which makes the integrals finite yet large. Therefore, to evaluate the zeros of the inverse Green's function $D_q^{-1}$, we need to consider the terms with the most divergent behavior. Under this constraint we reach
\begin{align}
    D_q^{-1} & \approx  \sum_{\bk} \frac{ f_{\bk}^2 }{2 d_{\bk}^3}\left[ \frac{\left(\tilde{q}_{x}\tilde{q}_y\right)^2}{16} \left(\tilde{\partial}_{xy}d_{\bk}\right)^2 - \omega^2 \right]. 
\end{align}
Recalling that $\tilde{q}_{x,y}= (q_x \pm q_y)$, we can see that the dispersion relation of the phason field up to some regularization-dependent coefficient is  
\begin{align}
    \omega \propto \left|(q_x - q_y)(q_x+q_y)\right|.
\end{align}
Note that similar to the results of the hot-spot model in the extreme limit of $\theta_{\delta}=0$, the dispersion relation could be a function of $(q_x \pm q_y)$ factors. Hence, at the low-doping limit, the anisotropic symmetry structure of the phason field becomes more pronounced. It should be also pointed out that the peculiar quadratic dispersion which arises at this level of the truncation of the Taylor expansion of $D_{q}^{-1}$, is due to the fact that in our model the manifold in which the band gap vanishes is a 1D nodal line instead of some isolated nodal points. Therefore, for more generic models, where the singularity of the correlation functions is a discrete zero-dimensional manifold, one should reproduce the more conventional linear behavior which is expected for phase modes.

\clearpage

\end{widetext}
\end{document}